\documentclass[aps,prb,twocolumn,showpacs,floats]{revtex4}
\usepackage{graphics}
\usepackage{amssymb}
\begin{document}

\title{Low temperature thermal transport at the interface of a topological insulator and a $d$-wave superconductor}
\author{Adam C. Durst}
\affiliation{Department of Physics and Astronomy, Hofstra University, Hempstead, NY 11549-0151}
\date{January 19, 2015}

\begin{abstract}
We consider the low-temperature thermal transport properties of the 2D proximity-induced superconducting state formed at the interface between a 3D strong topological insulator (TI) and a d-wave superconductor (dSC). This system is a playground for studying massless Dirac fermions, as they enter both as quasiparticles of the dSC and as surface states of the TI. For TI surface states with a single Dirac point, the four nodes in the interface-state quasiparticle excitation spectrum coalesce into a single node as the chemical potential, $\mu$, is tuned from above the impurity scattering rate ($|\mu| \gg \Gamma_{0}$) to below ($|\mu| \ll \Gamma_{0}$). We calculate, via Kubo formula, the universal limit ($T \rightarrow 0$) thermal conductivity, $\kappa_{0}$, as a function of $\mu$, as it is tuned through this transition. In the large and small $|\mu|$ limits, we obtain disorder-independent, closed-form expressions for $\kappa_{0}/T$. The large-$|\mu|$ expression is exactly half the value expected for a d-wave superconductor, a demonstration of the sense in which the TI surface topological metal is half of an ordinary 2D electron gas. Our numerical results for intermediate $|\mu|$ illustrate the nature of the transition between these limits, which is shown to depend on disorder in a well-defined manner.
\end{abstract}

\pacs{74.25.fc, 73.20.-r, 74.20.Rp, 74.45.+c}

\maketitle

\section{Introduction}
\label{sec:intro}
Topological insulators \cite{has10,qi11,moo10} (TI) represent a novel state of quantum matter that comes about due to the combined effects of spin-orbit interactions and time-reversal symmetry. \cite{kan05a,kan05b,fu07a,moo07,roy09}  Though characterized by a bulk band gap, they are adiabatically distinct from ordinary insulators and support protected gapless surface states.  In the case of a three-dimensional strong TI, these surface states form a novel two-dimensional topological metal with a spin-polarized massless Dirac energy spectrum.  The theoretical prediction and subsequent experimental discovery of TI states in 2D materials \cite{ber06,kon07} (HgTe/CdTe quantum wells), 3D materials \cite{fu07b,hsi08} (Bi$_x$Sb$_{1-x}$), and the cleaner, simpler, second generation 3D materials \cite{zha09,liu10,xia09} (Bi$_2$Se$_3$, Bi$_2$Te$_3$, and Sb$_2$Te$_3$) has led to great interest in this area, exploring both the fundamental physics as well as the potential for applications to fault-tolerant topological quantum computation \cite{fu08,nay08}.

The proximity of either magnetic materials or superconductors to the TI surface can induce an energy gap in the topological metal, resulting in even more exotic interface states. \cite{has10} Early on, Fu and Kane \cite{fu08} considered the proximity effect at the interface between a TI and a conventional $s$-wave superconductor, analyzing the proximity-induced superconducting interface state and finding that it should support Majorana bound states \cite{ali12,bee13,law09,sau10,hos11,akz14} at vortices.  Subsequent work has expanded this analysis in many directions, and has included the case of TIs proximity-coupled to unconventional superconductors of different pairing symmetries \cite{lin10a,lin10b,bla13}.  Such TI interface state superconductivity has been demonstrated, experimentally, both for the $s$-wave case \cite{sac11,vel12} and for the case of TIs coupled to high-$T_c$ cuprate $d$-wave superconductors \cite{zar12}.

It is this last case, that of the proximity-induced superconducting state at the interface of a 3D strong topological insulator (TI) and a $d$-wave superconductor (dSC), that is our focus here.  For simplicity, we will consider a TI with a surface state characterized by a single Dirac point at the origin of $k$-space, as is seen in the Bi$_2$Se$_3$ family of materials \cite{zha09,liu10}.  We are particularly interested in the low energy quasiparticle excitations of this interface state, a system in which massless Dirac fermions enter in two different ways, as both the surface states of the TI and the quasiparticles of the dSC.  For the former, the TI surface states, the massless Dirac fermions are isotropic, a consequence of band structure, and not pinned to the Fermi surface, such that one can tune through the Dirac point by varying the chemical potential.  They are described by a Dirac equation where the gamma matrices live in $2 \times 2$ spin space.  For the latter, the dSC quasiparticle states, the massless Dirac fermions are anisotropic, their energy spectrum squeezed in $k$-space, and are pinned to the Fermi surface at four nodal points.  They are described by a Dirac equation where the gamma matrices live in $2 \times 2$ particle-hole (Nambu) space.  The TI-dSC interface state that we consider mixes both spin and particle-hole space and will have quasiparticle excitations of its own, with features inherited from both of the above.

A useful probe for studying massless Dirac quasiparticles in $d$-wave superconductors has been low temperature thermal transport \cite{tai97,chi99,chi00,pro02,sut03,hil04,sun04,sut05,haw07,sun08}, measurements of which can be extrapolated to the particularly simple and interesting regime where temperature, $T$, is small compared to the impurity scattering rate, $\Gamma_0$.  This is known as the universal limit because thermal conductivity due to massless Dirac quasiparticles has been shown to be insensitive to disorder in this very low temperature regime. \cite{lee93,hir93,hir94,hir96,gra96,sen98,dur00}  In this paper, we examine the nature of the low energy quasiparticle excitations of the TI-dSC interface state by calculating the universal-limit thermal conductivity, $\kappa_0/T$, as a function of chemical potential, $\mu$.  Though the Hamiltonian for this interface state couples particle to hole and spin-up to spin-down, its quasiparticles carry a well-defined heat.  Thus, thermal transport tracks quasiparticle transport, and is therefore well-suited to probing the excitations of this system.  For $|\mu| \ll \Gamma_0$, it probes the single isotropic Dirac node inherited from the TI surface.  For $|\mu| \gg \Gamma_0$, it probes the four anisotropic Dirac nodes resulting from proximity-induced $d$-wave superconductivity.  We study both of these regimes and the transition between them as four nodes coalesce into one.

We begin in Sec.~\ref{sec:setup} by writing down the $4 \times 4$ Hamiltonian for the proximity-induced interface state, which mixes the spin-space Dirac equation of the TI surface with the particle-hole-space Dirac equation of the dSC, and then solve for the quasiparticle excitation spectrum.  In Sec.~\ref{sec:transport}, we calculate the matrix spectral function, derive the thermal current operator, and then use both of these to calculate the universal-limit thermal conductivity tensor via diagrammatic Kubo formula.  Closed form analytical expressions for $\kappa_0/T$ are obtained in both the large-$|\mu|$ and small-$|\mu|$ limits, both of which are discussed in Sec.~\ref{sec:analytical}.  Numerical results charting the disorder-dependent transition between these two limits are presented in Sec.~\ref{sec:numerical}.  Conclusions are discussed in Sec.~\ref{sec:conclusions}.

\section{Proximity-Induced Interface State}
\label{sec:setup}

\subsection{Hamiltonian}
\label{ssec:Hamiltonian}
We consider the proximity-induced superconducting state at the interface of a 3D strong topological insulator (TI) and a $d$-wave superconductor (dSC).  For a TI like those in the Bi$_2$Se$_3$ family, characterized by surface states with a single Dirac point at the $\Gamma$-point of the Brillouin zone, the TI surface state is described by the Hamiltonian \cite{fu08}
\begin{equation}
H_0 = \sum_k \psi_k^\dagger \left( v \vec{\bf \sigma} \cdot {\bf k} - \mu \right) \psi_k
\label{eq:H0}
\end{equation}
where $\psi_k = (c_{k\uparrow}, c_{k\downarrow})^T$ are electron annihilation operators, $v$ is the slope of the Dirac cone, $\mu$ is the chemical potential, $\vec{\bf \sigma}=(\sigma_1, \sigma_2)$ are spin Pauli matrices, and we have adopted units where $\hbar=1$.  Proximity to a dSC induces $d$-wave superconductivity and results in an interface state Hamiltonian \cite{fu08,lin10a,lin10b} that is most compactly expressed in the following $4 \times 4$ Nambu notation
\begin{equation}
H = \frac{1}{2}\sum_{k} \Psi_{k}^{\dagger} H_k \Psi_{k}
\label{eq:H}
\end{equation}
\begin{equation}
H_k = \left( v \vec{\bf \sigma} \cdot {\bf k} - \mu \right) \tau_3 + \Delta_k \tau_1
\label{eq:Hk}
\end{equation}
\begin{equation}
\Psi_{k}^{\dagger} = \left[ c_{k\uparrow}^{\dagger}, c_{k\downarrow}^{\dagger}, c_{-k\downarrow}, -c_{-k\uparrow} \right]
\label{eq:Psik}
\end{equation}
where the ${\bf \tau}$ are particle-hole Pauli matrices that mix the $\psi_k$ and $\psi_{-k}^\dagger$ blocks of $\Psi_k$, and the factor of 1/2 compensates for particle-hole double counting.  Here, the proximity-induced superconducting order parameter, $\Delta_k$, is of $d_{x^2-y^2}$ symmetry and is taken to be real.  (Note that in addition to this spin-singlet $d$-wave term, the form of the TI surface Hamiltonian allows for an additional, subdominant, spin-triplet (B$_{\rm 2u}$) $p$-wave term to also be induced via proximity to a dSC. \cite{bla13}  However, as shown by Linder {\it et al.\/} \cite{lin10a,lin10b}, a spin-triplet $p$-wave pairing amplitude in a TI only renormalizes the chemical potential and never gaps the surface energy spectrum.  Thus, while its inclusion here would likely result in a quantitative correction to the effect of the singlet term, it is not expected to change the essential physics.  Thus, for simplicity, we shall defer consideration of the triplet term to future work.)  Expanding Eq.~(\ref{eq:Hk}) by evaluating the outer products of the Pauli matrices yields the $4 \times 4$ Hamiltonian
\begin{equation}
H_k = \left[ \begin{array}{cccc} -\mu & vk^- & \Delta_k & 0 \\
vk^+ & -\mu & 0 & \Delta_k \\
\Delta_k & 0 & \mu & -vk^- \\
0 & \Delta_k & -vk^+ & \mu \end{array} \right]
\label{eq:Hk4x4}
\end{equation}
where $k^{\pm} \equiv k_x \pm ik_y$.

\subsection{Quasiparticle Excitation Spectrum}
\label{ssec:spectrum}
The quasiparticle excitation spectrum of the interface state is obtained by solving for the (positive) eigenvalues of $H_k$.  As shown in Ref.~\onlinecite{fu08} for the $s$-wave case, the resulting spectrum is
\begin{equation}
E_k = \sqrt{(\pm v|{\bf k}| - \mu)^2 + \Delta_k^2}
\label{eq:spectrum}
\end{equation}
Though the precise functional form of $\Delta_k$ is material-dependent, we can proceed, quite generally, as long as $\Delta_k$ satisfies two criteria: (1) It has $d_{x^2-y^2}$ symmetry and therefore changes sign along the lines $k_y = \pm k_x$.  (2) It vanishes faster than linearly with $k$ as $k \rightarrow 0$.  If these criteria are met, the quasiparticle spectrum will have the following properties.

For large $|\mu|$, there will be four nodal points in $k$-space, located at $\pm k_x = \pm k_y = \mu / \sqrt{2}v$, where one of the two branches in Eq.~(\ref{eq:spectrum}) goes to zero and quasiparticles can be excited for zero energy cost.  In the vicinity of each of these nodes,
\begin{equation}
E_k \approx \sqrt{v^2k_1^2 + v_\Delta^2k_2^2}
\label{eq:largemuspec}
\end{equation}
where $v_\Delta$ is the slope of $\Delta_k$ at the node and $k_1$ and $k_2$ define a local coordinate system, centered at each node, with the $k_1$-axis perpendicular to the local Fermi surface (pointing away from the origin of $k$-space) and the $k_2$-axis parallel to the local Fermi surface (pointing in the direction of increasing $\Delta_k$).  For energies small compared to $\mu$, the surfaces of constant energy are ellipses, elongated parallel to the local Fermi surface for $v > v_\Delta$ (as is typical in cuprate superconductors).  The presence of disorder smears out the nodes, exciting quasiparticles of energy less than or on the order of the impurity scattering rate, $\Gamma_0$.  For $T \ll \Gamma_0$, quasiparticle transport is dominated by these disorder-induced quasiparticles which reside within ellipses of semi-major axis $\Gamma_0/v_\Delta$ and semi-minor axis $\Gamma_0/v$ about each of the four nodes.

The nodes are distinct for $|\mu| \gg \Gamma_0$, but as $|\mu|$ decreases, the inter-node separation decreases, and for $|\mu| \ll \Gamma_0$, the nodes coalesce at the origin of $k$-space.  As long as $\Delta_k$ vanishes fast enough with decreasing $k$, as per condition (2) above, this transition reveals the underlying massless Dirac spectrum of the TI surface state.  Thus, for $|\mu| \ll \Gamma_0$,
\begin{equation}
E_k \approx v|{\bf k}|
\label{eq:smallmuspec}
\end{equation}
and the system thereby trades the four anisotropic nodes at nonzero ${\bf k}$ for a single isotropic node at the origin.  Note that this single node is, however, doubly degenerate, as it derives from both of the branches in Eq.~(\ref{eq:spectrum}).  For $|\mu|$ and $T$ small compared to $\Gamma_0$, quasiparticle transport is dominated by the disorder-induced quasiparticles that reside within the circle of radius $\Gamma_0/v$ about this isotropic node.

\section{Transport Calculation}
\label{sec:transport}
Following the approach employed in Refs.~\onlinecite{dur00} and \onlinecite{dur09}, we now proceed to calculate the universal-limit quasiparticle thermal conductivity for this system, as a function of chemical potential.  Key inputs to this calculation are the spectral function and thermal current operator, which we will calculate first and then utilize in our calculation of the thermal conductivity.

\subsection{Spectral Function}
\label{ssec:spectralfunction}
To obtain the spectral function, we begin by calculating the Matsubara Green's function.  Working in our 4-component Nambu basis, the $4 \times 4$ bare Green's function is obtained by inverting the Hamiltonian
\begin{equation}
G^0({\bf k},i\omega) = \left[ i\omega \openone - H_k \right]^{-1}
\label{eq:G0}
\end{equation}
where $H_k$ is the $4 \times 4$ Hamiltonian from Eq.~(\ref{eq:Hk4x4}).  The dressed Green's function is then found via Dyson's equation
\begin{equation}
G({\bf k},i\omega)^{-1} = G^0({\bf k},i\omega)^{-1} - \Sigma(i\omega)
\label{eq:Dyson}
\end{equation}
such that
\begin{equation}
G({\bf k},i\omega) = \left[ i\omega \openone -\Sigma(i\omega) - H_k \right]^{-1}
\label{eq:Gmatsu}
\end{equation}
where $\Sigma$ is the Matsubara self-energy matrix.  The retarded Green's function is then obtained by continuing $i\omega \rightarrow \omega + i\delta$
\begin{equation}
G^R({\bf k},\omega) = \left[ \omega \openone - \Sigma^R(\omega) - H_k \right]^{-1}
\label{eq:GR}
\end{equation}
where $\Sigma^R(\omega) = \Sigma(i\omega \rightarrow \omega + i\delta)$ is the retarded self-energy matrix.  We define the matrix spectral function, $A({\bf k},\omega)$, via
\begin{equation}
G({\bf k},i\omega) = \int_{\infty}^{\infty} d\omega^{\prime} \frac{A({\bf k},\omega^\prime)}{i\omega - \omega^\prime}
\label{eq:specfuncdef}
\end{equation}
such that
\begin{equation}
A({\bf k},\omega) = \frac{i}{2\pi} \left(G^R({\bf k},\omega) - G^A({\bf k},\omega) \right)
\label{eq:A}
\end{equation}
where $G^A = \left.G^{R}\right.^{\dagger}$ is the advanced Green's function.  Since our calculation will only require $A({\bf k},\omega \rightarrow 0)$, we need only calculate
\begin{equation}
G^R({\bf k},0) = \left[ \begin{array}{cccc} i\Gamma_0+\mu & -vk^- & -\Delta_k & 0 \\
-vk^+ & i\Gamma_0+\mu & 0 & -\Delta_k \\
-\Delta_k & 0 & i\Gamma_0-\mu & vk^- \\
0 & -\Delta_k & vk^+ & i\Gamma_0-\mu \end{array} \right]^{-1}
\label{eq:GR0}
\end{equation}
where $k^{\pm} \equiv k_x \pm ik_y$.  Here we have taken a simple form for the zero-frequency self-energy matrix, $\Sigma^R(\omega \rightarrow 0) = -i\Gamma_0 \openone$, where $\Gamma_0$ is a scalar constant, the impurity scattering rate.  In general, the full $4 \times 4$ self-energy matrix can be calculated for a particular disorder model, but this simple model captures the essential physics and establishes an energy scale for disorder.  Performing the inversion in Eq.~(\ref{eq:GR0}) yields the zero-frequency, matrix spectral function
\begin{equation}
A({\bf k},0) = \frac{\left( A_0 \openone_\sigma + A_1 \sigma_1 + A_2 \sigma_2 \right) \openone_\tau}{A_{\rm den}}
\label{eq:Ak0}
\end{equation}
where $\openone_\sigma$ is the intra-block (spin) $2 \times 2$ identity matrix, $\openone_\tau$ is the inter-block (particle-hole) $2 \times 2$ identity matrix, and
\begin{eqnarray}
\label{eq:A012den}
A_0 &=& \Gamma_0 \left( \Gamma_0^2 + \mu^2 + v^2 k^2 + \Delta_k^2 \right) \\
A_1 &=& 2 \Gamma_0 \mu v k_x \nonumber \\
A_2 &=& 2 \Gamma_0 \mu v k_y \nonumber \\
A_{\rm den} &=& \pi \left( \Gamma_0^2 + (vk-\mu)^2 + \Delta_k^2 \right) \left( \Gamma_0^2 + (-vk-\mu)^2 + \Delta_k^2 \right) \nonumber
\end{eqnarray}

\subsection{Thermal Current Operator}
\label{ssec:thermalcurrent}
To derive an expression for the thermal current density operator in this system, we generalize the approach developed for the $s$-wave superconductor case by Ambegaokar and Griffin \cite{amb65} and adapted for the $d$-wave superconductor case in Ref.~\onlinecite{dur00}.  We begin by expressing the Hamiltonian in terms of the coordinate-space field operators, $\psi_\uparrow({\bf x})$ and $\psi_\downarrow({\bf x})$, such that
\begin{eqnarray}
\label{eq:Hx}
H &=& H_0 + H_1 \\
H_0 &=& \int d^2x \begin{array}{c} ( \psi_\uparrow^\dagger, \psi_\downarrow^\dagger ) \\ \left. \right. \end{array}
\left( -iv \vec{\bf \sigma} \cdot {\bf \nabla} - \mu \right)
\left( \begin{array}{c} \psi_\uparrow \\ \psi_\downarrow \end{array} \right) \nonumber \\
H_1 &=& \frac{1}{2} \int\! d^2x \int\! d^2y\, \psi_{x\alpha}^\dagger \psi_{y\beta}^\dagger V({\bf x} - {\bf y}) \psi_{y\beta} \psi_{x\alpha} \nonumber
\end{eqnarray}
where $\alpha$ and $\beta$ are spin indices over which summation is implied, $V({\bf x} - {\bf y})$ is the effective potential that gives rise to the proximity-induced superconductivity, and we have adopted a compact notation whereby $\psi_\alpha \equiv \psi_{x\alpha} \equiv \psi_\alpha({\bf x})$ and $\psi_{y\beta} \equiv \psi_\beta({\bf y})$.  Performing the matrix multiplications, $H_0$ takes the form
\begin{eqnarray}
H_0 &=& \int d^2x \left[ -iv \left( \psi_\uparrow^\dagger \partial^- \psi_\downarrow + \psi_\downarrow^\dagger \partial^+ \psi_\uparrow \right)
-\mu \psi_\alpha^\dagger \psi_\alpha \right] \nonumber \\
&=& \int d^2x \left[ iv \left( \partial^- \psi_\uparrow^\dagger \psi_\downarrow + \partial^+ \psi_\downarrow^\dagger \psi_\uparrow \right)
-\mu \psi_\alpha^\dagger \psi_\alpha \right]
\label{eq:H0x1}
\end{eqnarray}
where $\partial^\pm \equiv \frac{\partial}{\partial x} \pm i\frac{\partial}{\partial y}$ and the second equality is the result of integration by parts.  Equations of motion for the field operators are obtained by noting that
\begin{equation}
i\dot{\psi}_\alpha = \left[ \psi_\alpha, H\right] \;\;\;\;\;\; i\dot{\psi}_\alpha^\dagger = \left[ \psi_\alpha^\dagger, H\right]
\label{eq:eqmotiondef}
\end{equation}
and applying fermion anticommutation relations.  Doing so, we find that
\begin{eqnarray}
\dot{\psi}_\uparrow &=& -v\partial^- \psi_\downarrow + i\varphi_x \psi_\uparrow \nonumber \\
\dot{\psi}_\downarrow &=& -v\partial^+ \psi_\uparrow + i\varphi_x \psi_\downarrow \nonumber \\
\dot{\psi}_\uparrow^\dagger &=& -v\partial^+ \psi_\downarrow^\dagger - i\psi_\uparrow^\dagger \varphi_x  \nonumber \\
\dot{\psi}_\downarrow^\dagger &=& -v\partial^- \psi_\uparrow^\dagger - i\psi_\downarrow^\dagger \varphi_x
\label{eq:eqmotion}
\end{eqnarray}
where we have defined
\begin{equation}
\varphi_x \equiv \varphi({\bf x}) \equiv \mu - \int d^2r\, V({\bf r}-{\bf x}) \psi_{r\gamma}^\dagger \psi_{r\gamma}
\label{eq:varphioperator}
\end{equation}

The thermal current density operator, ${\bf j}^\kappa({\bf x})$, is obtained via continuity with the thermal density operator, $h({\bf x})$.
\begin{equation}
\dot{h}({\bf x}) = - \nabla \cdot {\bf j}^\kappa({\bf x})
\label{eq:continuity}
\end{equation}
Since we have written our Hamiltonian such that all energies are measured with respect to the chemical potential, $h({\bf x})$ is equal to the Hamiltonian density operator and therefore defined via
\begin{equation}
H = \int d^2x\, h({\bf x})
\label{eq:hdef}
\end{equation}
and expressed as
\begin{widetext}
\begin{equation}
h({\bf x}) = -\frac{iv}{2} \left( \psi_\uparrow^\dagger \partial^- \psi_\downarrow - \partial^- \psi_\uparrow^\dagger \psi_\downarrow
+ \psi_\downarrow^\dagger \partial^+ \psi_\uparrow - \partial^+ \psi_\downarrow^\dagger \psi_\uparrow \right)
- \mu \psi_\alpha^\dagger \psi_\alpha + \frac{1}{2} \int\! d^2y\, \psi_{\alpha}^\dagger \psi_{y\beta}^\dagger V({\bf y} - {\bf x}) \psi_{y\beta} \psi_{\alpha}
\label{eq:hx}
\end{equation}
where we have taken $H_0$ to be the average of the first and second lines of Eq.~(\ref{eq:H0x1}).  Taking the time derivative and breaking the result into two pieces, we write
\begin{equation}
\dot{h}({\bf x}) = F_A + F_B
\label{eq:hdotdef}
\end{equation}
where
\begin{equation}
F_A \equiv -\frac{iv}{2} \left( \dot{\psi}_\uparrow^\dagger \partial^- \psi_\downarrow - \partial^- \dot{\psi}_\uparrow^\dagger \psi_\downarrow
+ \dot{\psi}_\downarrow^\dagger \partial^+ \psi_\uparrow - \partial^+ \dot{\psi}_\downarrow^\dagger \psi_\uparrow
+ \psi_\uparrow^\dagger \partial^- \dot{\psi}_\downarrow - \partial^- \psi_\uparrow^\dagger \dot{\psi}_\downarrow
+ \psi_\downarrow^\dagger \partial^+ \dot{\psi}_\uparrow - \partial^+ \psi_\downarrow^\dagger \dot{\psi}_\uparrow \right)
- \mu \dot{\psi}_\alpha^\dagger \psi_\alpha - \mu \psi_\alpha^\dagger \dot{\psi}_\alpha
\label{eq:FAdef}
\end{equation}
\begin{equation}
F_B \equiv \frac{1}{2} \int\! d^2y\, V({\bf y} - {\bf x}) \left( \dot{\psi}_{\alpha}^\dagger \psi_{y\beta}^\dagger \psi_{y\beta} \psi_{\alpha}
+ \psi_{\alpha}^\dagger \dot{\psi}_{y\beta}^\dagger \psi_{y\beta} \psi_{\alpha}
+ \psi_{\alpha}^\dagger \psi_{y\beta}^\dagger \dot{\psi}_{y\beta} \psi_{\alpha}
+ \psi_{\alpha}^\dagger \psi_{y\beta}^\dagger \psi_{y\beta} \dot{\psi}_{\alpha} \right)
\label{eq:FBdef}
\end{equation}
The first piece, $F_A$, can be reorganized by using the equations of motion (\ref{eq:eqmotiondef}) to sub in for the dotted field operators, regrouping terms, and then applying the equations of motion again.  Doing so, we find that
\begin{equation}
F_A = \frac{iv}{2} \left[ \partial^- ( \dot{\psi}_\uparrow^\dagger \psi_\downarrow ) + \partial^+ ( \dot{\psi}_\downarrow^\dagger \psi_\uparrow )
- \partial^- ( \psi_\uparrow^\dagger \dot{\psi}_\downarrow ) - \partial^+ ( \psi_\downarrow^\dagger \dot{\psi}_\uparrow ) \right]
- \int\! d^2y\, V({\bf y} - {\bf x}) \left( \dot{\psi}_{\alpha}^\dagger \psi_{y\beta}^\dagger \psi_{y\beta} \psi_{\alpha}
+ \psi_{\alpha}^\dagger \psi_{y\beta}^\dagger \psi_{y\beta} \dot{\psi}_{\alpha} \right)
\label{eq:FA}
\end{equation}
Combining this with $F_B$ and applying the continuity equation (\ref{eq:continuity}), we see that it is natural to write the thermal current density operator as the sum of two terms
\begin{equation}
{\bf j}^\kappa \equiv {\bf u}_1 + {\bf u}_2
\label{eq:u1u2def}
\end{equation}
where
\begin{equation}
\nabla \cdot {\bf u}_1 = - \frac{iv}{2} \left[ \partial^- ( \dot{\psi}_\uparrow^\dagger \psi_\downarrow ) + \partial^+ ( \dot{\psi}_\downarrow^\dagger \psi_\uparrow )
- \partial^- ( \psi_\uparrow^\dagger \dot{\psi}_\downarrow ) - \partial^+ ( \psi_\downarrow^\dagger \dot{\psi}_\uparrow ) \right]
\label{eq:divu1}
\end{equation}
\begin{equation}
\nabla \cdot {\bf u}_2 = \frac{1}{2} \int\! d^2y\, V({\bf y} - {\bf x}) \left[ \left( \dot{\psi}_{\alpha}^\dagger \psi_{y\beta}^\dagger \psi_{y\beta} \psi_{\alpha}
+ \psi_{\alpha}^\dagger \psi_{y\beta}^\dagger \psi_{y\beta} \dot{\psi}_{\alpha} \right)
- \left( \psi_{\alpha}^\dagger \dot{\psi}_{y\beta}^\dagger \psi_{y\beta} \psi_{\alpha}
+ \psi_{\alpha}^\dagger \psi_{y\beta}^\dagger \dot{\psi}_{y\beta} \psi_{\alpha} \right) \right]
\label{eq:divu2}
\end{equation}

Expansion of the $\partial^\pm$ operators reveals that the right-hand-side of Eq.~(\ref{eq:divu1}) is easily expressed as a divergence.  Doing so, we extract
\begin{equation}
{\bf u}_1 = -\frac{iv}{2} \left[ \left( \left( \dot{\psi}_\uparrow^\dagger \psi_\downarrow + \dot{\psi}_\downarrow^\dagger \psi_\uparrow \right) \hat{x}
-i \left( \dot{\psi}_\uparrow^\dagger \psi_\downarrow - \dot{\psi}_\downarrow^\dagger \psi_\uparrow \right) \hat{y} \right)
- \left( \left( \psi_\uparrow^\dagger \dot{\psi}_\downarrow + \psi_\downarrow^\dagger \dot{\psi}_\uparrow \right) \hat{x}
-i \left( \psi_\uparrow^\dagger \dot{\psi}_\downarrow - \psi_\downarrow^\dagger \dot{\psi}_\uparrow \right) \hat{y} \right) \right]
\label{eq:u1psi}
\end{equation}
which, in $4 \times 4$ Nambu notation, becomes
\begin{equation}
{\bf u}_1({\bf x},t) = -\frac{iv}{4} \left[ \dot{\Psi}^\dagger \vec{\sigma} \tau_3 \Psi
- \Psi^\dagger \vec{\sigma} \tau_3 \dot{\Psi} \right]
\label{eq:u1nambu}
\end{equation}
where $\Psi^\dagger = \Psi^\dagger({\bf x},t)=[ \psi_\uparrow^\dagger, \psi_\downarrow^\dagger, \psi_\downarrow, - \psi_\uparrow]$ and $\vec{\sigma} = \sigma_1 \hat{x} + \sigma_2 \hat{y}$.  Fourier transforming in space and time yields
\begin{equation}
{\bf u}_1({\bf q},\Omega) = \frac{1}{2} \sum_{k\omega} \Psi_k^\dagger \left( \omega + \frac{\Omega}{2} \right) v \vec{\sigma} \tau_3 \Psi_{k+q}
\label{eq:u1q}
\end{equation}
where we have used the shorthand $\Psi_k \equiv \Psi({\bf k},\omega)$ and $\Psi_{k+q} \equiv \Psi({\bf k+q},\omega+\Omega)$.

To obtain ${\bf u}_2$, we take the space-time Fourier transform of Eq.~(\ref{eq:divu2}).  Doing so yields
\begin{equation}
i{\bf q} \cdot {\bf u}_2({\bf q},\Omega) = \frac{1}{2} \int\! d^2x\, d^2y\, dt\, V({\bf y} - {\bf x})
\left( e^{-i{\bf q} \cdot {\bf x}} - e^{-i{\bf q} \cdot {\bf y}} \right)
\left( \dot{\psi}_{x\alpha}^\dagger \psi_{y\beta}^\dagger \psi_{y\beta} \psi_{x\alpha}
+ \psi_{x\alpha}^\dagger \psi_{y\beta}^\dagger \psi_{y\beta} \dot{\psi}_{x\alpha} \right)
= X_1 + X_2 - Y_1 - Y_2
\label{eq:divu2q}
\end{equation}
where we have labeled each of the four resulting terms: $X_1$, $X_2$, $Y_1$, and $Y_2$.  Inserting a Fourier representation for the potential and each of the field operators, the $X_1$ term takes the form
\begin{equation}
X_1 = \frac{i}{2} \sum_{k_1...k_5} \sum_{\omega_1...\omega_4} \omega_1 V_{k_5} c_{k_1\alpha}^\dagger c_{k_2\beta}^\dagger c_{k_3\beta} c_{k_4\alpha}\,
\delta(k_4 - k_1 - k_5 - q) \delta(k_3 - k_2 + k_5) \delta(\omega_1 + \omega_2 - \omega_3 - \omega_4 + \Omega)
\label{eq:X1a}
\end{equation}
\end{widetext}
Making a mean field approximation, retaining only the terms for which the average values are over $({\bf k}\uparrow,-{\bf k}\downarrow)$ pairs (reduced approximation), and noting that $\langle c_{k\uparrow}^\dagger c_{-k\downarrow}^\dagger \rangle$ is an even function of $\omega$, this becomes
\begin{equation}
X_1 = i\sum_{k\omega} (\omega - \Omega) \Delta_k^* c_{k-q\uparrow}^\dagger c_{-k\downarrow}^\dagger
\label{eq:X1b}
\end{equation}
where
\begin{equation}
\Delta_k \equiv \sum_{k^\prime \omega^\prime} V_{k-k^\prime} \langle c_{k^\prime\uparrow}^\dagger c_{-k^\prime\downarrow}^\dagger \rangle
\label{eq:Deltak}
\end{equation}
is the superconducting order parameter for the interface state.  Repeating this calculation for $X_2$, $Y_1$, and $Y_2$, and taking $\Delta_k$ to be real, we find that
\begin{eqnarray}
\lefteqn{{\bf q} \cdot {\bf u}_2({\bf q},\Omega) = \sum_{k\omega} (\Delta_{k+q} - \Delta_k)} \nonumber \\
&& \times \left[ \omega c_{k\uparrow}^\dagger c_{-(k+q)\downarrow}^\dagger + (\omega + \Omega) c_{-k\downarrow} c_{k+q\uparrow} \right]
\label{eq:qdotu2}
\end{eqnarray}
In the $q \rightarrow 0$ limit,
\begin{equation}
\Delta_{k+q} - \Delta_k = {\bf q} \cdot \frac{\partial \Delta_k}{\partial {\bf k}} = {\bf q} \cdot {\bf v}_{\Delta k}
\label{eq:vDeltadef}
\end{equation}
where ${\bf v}_{\Delta k}$ is the slope of the order parameter (the gap velocity) at $k$.  Plugging into Eq.~(\ref{eq:qdotu2}) and taking the $\Omega \rightarrow 0$ limit, we find that
\begin{equation}
{\bf u}_2(0,0) = \sum_{k\omega} \left( \omega + \frac{\Omega}{2} \right) {\bf v}_{\Delta k} \left( c_{k\uparrow}^\dagger c_{-(k+q)\downarrow}^\dagger + c_{-k\downarrow} c_{k+q\uparrow} \right)
\label{eq:u2scalar}
\end{equation}
which, in the $4 \times 4$ Nambu notation, becomes
\begin{equation}
{\bf u}_2(0,0) = \frac{1}{2} \sum_{k\omega} \Psi_k^\dagger \left( \omega + \frac{\Omega}{2} \right) \vec{\bf v}_{\Delta k}\, \tau_1\, \Psi_{k+q}
\label{eq:u2nambu}
\end{equation}
Thus, in the $q,\Omega \rightarrow 0$ limit (which is the limit where we will need it), the thermal current density operator is
\begin{equation}
{\bf j}^\kappa(0,0) = \frac{1}{2} \sum_{k\omega} \Psi_k^\dagger \left( \omega + \frac{\Omega}{2} \right) {\bf v}_M \Psi_{k+q}
\label{eq:jkappa}
\end{equation}
where
\begin{equation}
\vec{\bf v}_M \equiv v \vec{\sigma} \tau_3 + \vec{\bf v}_{\Delta k}\, \tau_1
\label{eq:vM}
\end{equation}
is a vector in coordinate space and a matrix in our $4 \times 4$ Nambu space.  Here, the first term derives from the massless Dirac spectrum of the TI surface and has inherited the interesting spin structure thereof, while the second term derives from the the $d$-wave order parameter of the proximity-induced superconductivity.

\subsection{Thermal Conductivity}
\label{ssec:thermalcond}
\begin{figure}
\centerline{\resizebox{2in}{!}{\includegraphics{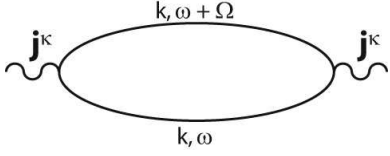}}}
\caption{Feynman diagram representing the bare bubble thermal current-current correlation function, $\tensor{\Pi}_\kappa(i\Omega)$.  On each vertex sits a thermal current density operator, ${\bf j}^\kappa$.  Each propagator line denotes a Green's function dressed with disorder self-energy, $G({\bf k},i\omega)$ and $G({\bf k},i\omega+i\Omega)$.}
\label{fig:bubble}
\end{figure}

With the spectral function and thermal current density operator in hand, we can proceed to calculate the thermal conductivity in the zero-temperature, zero-frequency limit.  For $d$-wave superconductors, this limit is known as the universal limit because thermal conductivity has been shown to be insensitive to disorder in this regime. \cite{lee93,hir93,hir94,hir96,gra96,sen98,dur00}  We can calculate the thermal conductivity tensor, $\tensor{\kappa}(T)$, for the case at hand by appealing to the fluctuation-dissipation theorem as expressed in the Kubo formula \cite{mah90}
\begin{equation}
\frac{\tensor{\kappa}(T)}{T} = -\lim_{\Omega \rightarrow 0} \frac{\mbox{Im}\,\tensor{\Pi}_\kappa^R(\Omega)}{T^2 \Omega}
\label{eq:kubo}
\end{equation}
where we obtain the retarded current-current correlation function via analytic continuation from the Matsubara function.
\begin{equation}
\tensor{\Pi}_\kappa^R(\Omega) = \tensor{\Pi}_\kappa(i\Omega \rightarrow \Omega + i\delta)
\label{eq:Picontinue}
\end{equation}
For simplicity, we proceed by calculating the bare bubble Feynman diagram shown in Fig.~\ref{fig:bubble}, noting that vertex corrections have been shown to be small for the $d$-wave superconductor case \cite{dur00} and deferring to future work their calculation for the case at hand.  Doing so, the Matsubara thermal current-current correlation function takes the form
\begin{eqnarray}
\tensor{\Pi}_\kappa (i\Omega) &=& \frac{1}{2} \frac{1}{\beta} \sum_{i\omega} \sum_k (i\omega + \frac{i\Omega}{2})^2 \nonumber \\
&\times& \mbox{Tr} \left[ G({\bf k},i\omega) {\bf v}_M G({\bf k},i\omega+i\Omega) {\bf v}_M \right]
\label{eq:Pidef}
\end{eqnarray}
where $\beta = 1/k_BT$, the $\omega$-sum is over fermionic Matsubara frequencies, the $k$-sum is over the Brillouin zone, the trace is over Nambu space, and the factor of $1/2$ out front compensates for the particle-hole double counting that is inherent in our $4 \times 4$ Nambu formalism.  Inserting a matrix spectral representation, as defined in Eqs.~(\ref{eq:specfuncdef}) and (\ref{eq:A}), for each of the Green's functions, this becomes
\begin{eqnarray}
\tensor{\Pi}_\kappa (i\Omega) &=& \frac{1}{2} \sum_k \int d\omega_1\, d\omega_2\, S(i\Omega) \nonumber \\
&\times& \mbox{Tr} \left[ A({\bf k},\omega_1) {\bf v}_M A({\bf k},\omega_2) {\bf v}_M \right]
\label{eq:Pimat}
\end{eqnarray}
where
\begin{equation}
S(i\Omega) = \frac{1}{\beta} \sum_{i\omega} (i\omega+\frac{i\Omega}{2})^2 \frac{1}{i\omega - \omega_1} \frac{1}{i\omega + i\Omega - \omega_2}
\label{eq:Smat}
\end{equation}
Evaluating the Matsubara sum via contour integration (see Refs.~\onlinecite{amb65} and \onlinecite{dur00} for a discussion of the technical points) and continuing $i\Omega \rightarrow \Omega + i\delta$, we obtain the retarded function
\begin{equation}
S_R(\Omega) = \frac{(\omega_1 + \Omega/2)^2 n_F(\omega_1) - (\omega_2 - \Omega/2)^2 n_F(\omega_2)}{\omega_1 - \omega_2 + \Omega + i\delta}
\label{eq:Sret}
\end{equation}
where $n_F(\omega)=1/(e^{\beta\omega}+1)$ is the Fermi function.  Since the retarded and advanced Green's functions are hermitian conjugates, the spectral function defined in Eq.~(\ref{eq:A}) must be hermitian
\begin{equation}
A^\dagger = -i\frac{\left. G^R \right.^\dagger - \left. G^A \right.^\dagger}{2\pi} = -i\frac{G^A - G^R}{2\pi} = i\frac{G^R - G^A}{2\pi} = A
\label{eq:Adagger}
\end{equation}
And since ${\bf v}_M$ is also hermitian, the trace in Eq.~(\ref{eq:Pimat}) must be real
\begin{eqnarray}
\lefteqn{\mbox{Tr} \left[ A_1 {\bf v}_M A_2 {\bf v}_M \right]^* = \mbox{Tr} \left[ \left( A_1 {\bf v}_M A_2 {\bf v}_M \right)^T \right]^*} \nonumber \\
&=& \mbox{Tr} \left[ \left( A_1 {\bf v}_M A_2 {\bf v}_M \right)^\dagger \right] = \mbox{Tr} \left[ {\bf v}_M^\dagger A_2^\dagger {\bf v}_M^\dagger A_1^\dagger \right] \nonumber \\
&=& \mbox{Tr} \left[ {\bf v}_M A_2 {\bf v}_M A_1 \right] = \mbox{Tr} \left[ A_1 {\bf v}_M A_2 {\bf v}_M \right]
\label{eq:TrReal}
\end{eqnarray}
Therefore
\begin{eqnarray}
\mbox{Im}\, \tensor{\Pi}_\kappa^R(\Omega) &=& \frac{1}{2} \sum_k \int d\omega_1\ d\omega_2\ \mbox{Im} S_R(\Omega) \nonumber \\
&\times& \mbox{Tr} \left[ A({\bf k},\omega_1) {\bf v}_M A({\bf k},\omega_2) {\bf v}_M \right]
\label{eq:ImPiRet}
\end{eqnarray}
where
\begin{eqnarray}
\mbox{Im} S_R(\Omega) &=& \pi (\omega_1 + \frac{\Omega}{2})^2 \left( n_F(\omega_1 + \Omega) - n_F(\omega_1) \right) \nonumber \\
&\times& \delta(\omega_1 + \Omega - \omega_2)
\label{eq:ImSret}
\end{eqnarray}
Plugging into Eq.~(\ref{eq:kubo}) and taking the $\Omega \rightarrow 0$ limit yields an expression for the thermal conductivity tensor.
\begin{eqnarray}
\frac{\tensor{\kappa}(T)}{T} &=& \frac{\pi}{2} \int d\omega \left( \frac{\omega}{T} \right)^2 \left( - \frac{\partial n_F}{\partial \omega} \right) \nonumber \\
&\times& \sum_k \mbox{Tr} \left[ A({\bf k},\omega) {\bf v}_M A({\bf k},\omega) {\bf v}_M  \right]
\label{eq:kappaT}
\end{eqnarray}
In the zero temperature limit, $(\omega/T)^2 (-\partial n_F / \partial \omega)$ is sharply peaked at $\omega=0$.  Thus, evaluating the integral
\begin{equation}
\int_{-\infty}^{\infty} \! d\omega \left( \frac{\omega}{T} \right)^2 \left(-\frac{\partial n_F}{\partial \omega}\right) = \frac{\pi^2 k_B^2}{3}
\label{eq:integ1}
\end{equation}
we find that the universal limit thermal conductivity tensor takes the form
\begin{equation}
\frac{\tensor{\kappa}_0}{T} \equiv \left. \frac{\tensor{\kappa}(T)}{T} \right|_{T \rightarrow 0} = \frac{\pi^3 k_B^2}{6} \sum_k
\mbox{Tr} \left[ A({\bf k},0) {\bf v}_M A({\bf k},0) {\bf v}_M  \right]
\label{eq:kappa0a}
\end{equation}
where $A({\bf k},0)$ is the spectral function that we evaluated in Eqs.~(\ref{eq:Ak0}) and (\ref{eq:A012den}).

Introducing the shorthand Tr$AvAv$ for the trace in the above expression and plugging in for $A({\bf k},0)$ via Eq.~(\ref{eq:Ak0}) and for ${\bf v}_M$ via Eq.~(\ref{eq:vM}) yields
\begin{eqnarray}
\lefteqn{\mbox{Tr}AvAv = \frac{1}{A_{\rm den}^2} \mbox{Tr} \left[ \left( N \openone_\tau
\left( v \vec{\sigma} \tau_3 + \vec{\bf v}_{\Delta k}\, \openone_\sigma \tau_1 \right) \right)^2 \right]} \nonumber \\
&=& \frac{1}{A_{\rm den}^2} \mbox{Tr} \left[ v^2 (N \vec{\sigma})^2 \openone_\tau + \vec{\bf v}_{\Delta k} \vec{\bf v}_{\Delta k} N^2 \openone_\tau \right]
\label{eq:TrAvAv1}
\end{eqnarray}
where $N \equiv A_0 \openone_\sigma + A_1 \sigma_1 + A_2 \sigma_2$ and we have made use of the multiplicative properties of the particle-hole ($\tau$) Pauli matrices.  Noting that $\vec{\sigma} = \sigma_1 \hat{x} + \sigma_2 \hat{y}$, making use of the multiplicative properties of the spin ($\sigma$) Pauli matrices, evaluating the trace, and plugging back into Eq.~(\ref{eq:kappa0a}), we find that
\begin{eqnarray}
\frac{\tensor{\kappa}_0}{T} &=& 4 \frac{\pi^3 k_B^2}{6} \biggl[ v^2 (\hat{x}\hat{x}+\hat{y}\hat{y}) \sum_k \frac{A_0^2}{A_{\rm den}^2} \nonumber \\
&+& v^2 (\hat{x}\hat{x}-\hat{y}\hat{y}) \sum_k \frac{A_1^2-A_2^2}{A_{\rm den}^2} \nonumber \\
&+& v^2 (\hat{x}\hat{y}+\hat{y}\hat{x}) \sum_k \frac{2A_1 A_2}{A_{\rm den}^2} \nonumber \\
&+& \sum_k \vec{\bf v}_{\Delta k} \vec{\bf v}_{\Delta k} \frac{A_0^2 + A_1^2 + A_2^2}{A_{\rm den}^2} \biggr]
\label{eq:kappa0b}
\end{eqnarray}
Since $\Delta_k$ is of $d_{x^2-y^2}$ symmetry, it must be an even function of both $k_x$ and $k_y$.  Therefore, as defined in Eq.~(\ref{eq:A012den}), $A_0$ and $A_{\rm den}$ are even functions of $k_x$ and $k_y$ while $A_1$ is odd in $k_x$ but even in $k_y$ and $A_2$ is even in $k_x$ but odd in $k_y$.  As a result
\begin{equation}
\sum_k \frac{2A_1 A_2}{A_{\rm den}^2} = \int \frac{dk_x}{2\pi} \int \frac{dk_y}{2\pi} \frac{2A_1 A_2}{A_{\rm den}^2} = 0
\label{eq:sum0a}
\end{equation}
And since exchange of $k_x$ for $k_y$ sends $\Delta_k$ to $-\Delta_k$, it leaves $A_{\rm den}$ invariant but exchanges $A_1$ for $A_2$.  Therefore
\begin{equation}
\sum_k \frac{A_1^2}{A_{\rm den}^2} = \sum_k \frac{A_2^2}{A_{\rm den}^2}
\label{eq:sum0b}
\end{equation}
Thus, only the first and fourth terms in Eq.~(\ref{eq:kappa0b}) survive.  Noting that $\hat{x}\hat{x}+\hat{y}\hat{y}$ is just the identity tensor, $\tensor{\openone}$, plugging in for $A_0$, $A_1$, $A_2$, and $A_{\rm den}$ from Eq.~(\ref{eq:A012den}), and restoring $\hbar$ in the prefactor, we obtain the following expression for the thermal conductivity in the zero temperature limit
\begin{equation}
\frac{\tensor{\kappa}_0}{T} = \frac{k_B^2}{3\hbar} 2\pi^3 \left[ v^2 \tensor{\openone} \sum_k P_k
+ \sum_k \vec{\bf v}_{\Delta k} \vec{\bf v}_{\Delta k} \left( P_k + Q_k \right) \right]
\label{eq:kappa0c}
\end{equation}
where $P_k \equiv A_0^2 / A_{\rm den}^2$ and $Q_k \equiv (A_1^2 + A_2^2) / A_{\rm den}^2$ take the form
\begin{equation}
P_k = \frac{1}{4} \biggl[ \frac{\Gamma_0/\pi}{\Gamma_0^2 + (vk-\mu)^2 + \Delta_k^2} + \frac{\Gamma_0/\pi}{\Gamma_0^2 + (-vk-\mu)^2 + \Delta_k^2} \biggr]^2
\label{eq:Pk}
\end{equation}
\begin{equation}
Q_k = \frac{1}{4} \biggl[ \frac{\Gamma_0/\pi}{\Gamma_0^2 + (vk-\mu)^2 + \Delta_k^2} - \frac{\Gamma_0/\pi}{\Gamma_0^2 + (-vk-\mu)^2 + \Delta_k^2} \biggr]^2
\label{eq:Qk}
\end{equation}
Note that this result depends on integrals of the squares of sums and differences of Lorentzians centered about the zeros of the two branches of the quasiparticle excitation spectrum, Eq.~(\ref{eq:spectrum}), of width given by the impurity scattering rate.  For $\mu \gg \Gamma_0$, $\tensor{\kappa_0}$ is dominated by impurity-induced quasiparticles in the vicinity of the zeros of the $(+)$ branch.  For $\mu \ll -\Gamma_0$, the $(-)$ branch dominates.  For $|\mu| \ll \Gamma_0$, both branches contribute.

\section{Analytical Results}
\label{sec:analytical}
In both the large-$|\mu|$ and small-$|\mu|$ limits ($|\mu| \gg \Gamma_0$ and $|\mu| \ll \Gamma_0$) the quasiparticle excitation spectrum simplifies, as described in Sec.~\ref{ssec:spectrum}, and can be linearized about nodal points in $k$-space.  As a result, in these limits, we can obtain simple, closed-form expressions for the zero-temperature thermal conductivity.  This is shown in the following sections.

\subsection{Large-$|\mu|$ Limit}
\label{ssec:largemu}
For $\mu \gg \Gamma_0$, the Lorentzians in Eqs.~(\ref{eq:Pk}) and (\ref{eq:Qk}) are sharply peaked about four nodal points, located at $\pm k_x = \pm k_y = \mu/\sqrt{2}v$, and are well separated from each other.  We can therefore replace the $k$-sum in Eq.~(\ref{eq:kappa0c}) by the sum of four integrals over local scaled coordinates, $p_1$ and $p_2$, defined about the nodal points
\begin{equation}
\sum_k \rightarrow \sum_{j=1}^4 \int \frac{d^2k}{(2\pi)^2} \rightarrow \sum_{j=1}^4 \int \frac{d^2p}{(2\pi)^2 v v_\Delta}
\label{eq:4nodes}
\end{equation}
where the integrals can be extended to infinity because the integrands are so sharply peaked about each node.  Here, $p_1 \equiv v k_1$ and $p_2 \equiv v_{\Delta} k_2$, and at each node, $\hat{k}_1$ and $\hat{k}_2$ point, respectively, perpendicular to and parallel to the local Fermi surface, with $\hat{k}_2$ in the direction of increasing $\Delta_k$.  In terms of these scaled coordinates, $\Delta_k \approx p_2$ and $vk = \mu + p_1$, so $vk-\mu = p_1$ and $-vk-\mu = -(2\mu+p_1) \approx -2\mu$.  Therefore, since $\mu \gg \Gamma_0$, the second Lorentzian can be neglected with respect to the first in both Eq.~(\ref{eq:Pk}) and Eq.~(\ref{eq:Qk}) and we find that
\begin{equation}
P_k = Q_k = \frac{1}{4} \left( \frac{\Gamma_0/\pi}{\Gamma_0^2 + p^2} \right)^2
\label{eq:PkQklargemu}
\end{equation}
where $p \equiv \sqrt{p_1^2 + p_2^2}$.  Evaluating the integral
\begin{equation}
\int \frac{d^2p}{(2\pi)^2} \left( \frac{\Gamma_0/\pi}{\Gamma_0^2 + p^2} \right)^2 = \frac{1}{4\pi^3}
\label{eq:integ2}
\end{equation}
and noting that the sum over nodes of the outer product of $\vec{\bf v}_{\Delta k}$ with itself at each node is
\begin{equation}
\sum_{j=1}^4 \vec{\bf v}_\Delta^{(j)} \vec{\bf v}_\Delta^{(j)} = 2 v_\Delta^2 \tensor{\openone}
\label{eq:vDvD}
\end{equation}
we find that
\begin{equation}
\sum_k P_k = 4 \cdot \frac{1}{v v_\Delta} \cdot \frac{1}{4} \cdot \frac{1}{4\pi^3} = \frac{1}{4\pi^3 v v_\Delta}
\label{eq:sumPk}
\end{equation}
\begin{equation}
\sum_k \vec{\bf v}_{\Delta k} \vec{\bf v}_{\Delta k} \left( P_k + Q_k \right)
= 2 v_\Delta^2 \tensor{\openone} \cdot \frac{1}{v v_\Delta} \cdot 2 \cdot \frac{1}{4} \cdot \frac{1}{4\pi^3}
= \frac{v_\Delta^2 \tensor{\openone}}{4\pi^3 v v_\Delta}
\label{eq:sumPkQk}
\end{equation}
Therefore, the thermal conductivity tensor reduces to a scalar, $\tensor{\kappa}_0 = \kappa_0 \tensor{\openone}$, with the simple form
\begin{equation}
\frac{\kappa_0}{T} = \frac{1}{2} \frac{k_B^2}{3\hbar} \left( \frac{v}{v_\Delta} + \frac{v_\Delta}{v} \right)
\label{eq:kappa0largemu}
\end{equation}
The same result is obtained for $\mu \ll -\Gamma_0$, where it is the first Lorentzian in Eqs.~(\ref{eq:Pk}) and (\ref{eq:Qk}) that can be neglected with respect to the second.  Note that this expression is independent of disorder and is only a function of the velocity anisotropy, $v/v_\Delta$, which depends on both $\mu$ and material parameters.  Note also that this is exactly half the value obtained (per layer) for the case of an ordinary $d$-wave superconductor. \cite{dur00}  This is because, unlike the $d$-wave superconductor case where the electron dispersion is spin-degenerate, here the TI surface state is nondegenerate and only one of the two branches of the quasiparticle excitation spectrum (Eq.~(\ref{eq:spectrum})) contributes to the thermal conductivity.  For $\mu \gg \Gamma_0$, the $(+)$ branch (first Lorentzian) contributes.  For $\mu \ll -\Gamma_0$, the $(-)$ branch (second Lorentzian) contributes.  This factor of two is a clear and measurable demonstration of the sense in which the TI surface topological metal is ``half'' of an ordinary 2D electron gas \cite{fu08}.

\subsection{Small-$|\mu|$ Limit}
\label{ssec:smallmu}
For $|\mu| \ll \Gamma_0$, the four anisotropic nodes of the prior section have coalesced into a single isotropic node at the origin of $k$-space.  The first and second Lorentzians in Eqs.~(\ref{eq:Pk}) and (\ref{eq:Qk}) are approximately equal and peaked at the origin.  The $k$-sum in Eq.~(\ref{eq:kappa0c}) can be replaced by a single integral about scaled coordinates, $p_1=vk_x$ and $p_2=vk_y$, and extended to infinity.
\begin{equation}
\sum_k \rightarrow \int \frac{d^2k}{(2\pi)^2} \rightarrow \int \frac{d^2p}{(2\pi)^2 v^2}
\label{eq:1node}
\end{equation}
In these scaled coordinates, $vk = p = \sqrt{p_1^2+p_2^2}$, and since $|\mu| \ll \Gamma_0$, $vk-\mu \approx p$ and $-vk-\mu \approx -p$.  As long as $\Delta_k$ vanishes fast enough with decreasing $k$, as per condition (2) of Sec.~\ref{ssec:spectrum}, $\Delta_k^2$ and $\vec{\bf v}_{\Delta k} \vec{\bf v}_{\Delta k}$ can be neglected in Eqs.~(\ref{eq:kappa0c}-\ref{eq:Qk}) compared to larger terms.  As a result, the two Lorentzians add in $P_k$ and cancel out in $Q_k$.
\begin{equation}
P_k = \left( \frac{\Gamma_0/\pi}{\Gamma_0^2 + p^2} \right)^2 \;\;\;\;\;\; Q_k \approx 0
\label{eq:PkQksmallmu}
\end{equation}
Once again making use of the integral in Eq.~(\ref{eq:integ2}), we find that
\begin{equation}
\sum_k P_k = \frac{1}{v^2} \int \frac{d^2p}{(2\pi)^2} \left( \frac{\Gamma_0/\pi}{\Gamma_0^2 + p^2} \right)^2 = \frac{1}{4\pi^3 v^2}
\label{eq:sumPksmallmu}
\end{equation}
\begin{equation}
\sum_k \vec{\bf v}_{\Delta k} \vec{\bf v}_{\Delta k} \left( P_k + Q_k \right) \approx 0
\label{eq:sumPkQksmallmu}
\end{equation}
Therefore, the thermal conductivity tensor again reduces to a scalar, now with an even simpler form
\begin{equation}
\frac{\kappa_0}{T} = \frac{k_B^2}{3\hbar} \frac{1}{2}
\label{eq:kappa0smallmu}
\end{equation}
Here, both branches of the quasiparticle spectrum have contributed to the thermal conductivity, and one obtains precisely the result one would expect for a single isotropic massless Dirac node.  This expression is clearly independent of disorder and is just the standard $d$-wave superconductor result \cite{dur00} for an anisotropy ratio of one, divided by a factor of four since there is only one node here rather than four.

\section{Numerical Results}
\label{sec:numerical}
\begin{figure}
\centerline{\resizebox{3.25in}{!}{\includegraphics{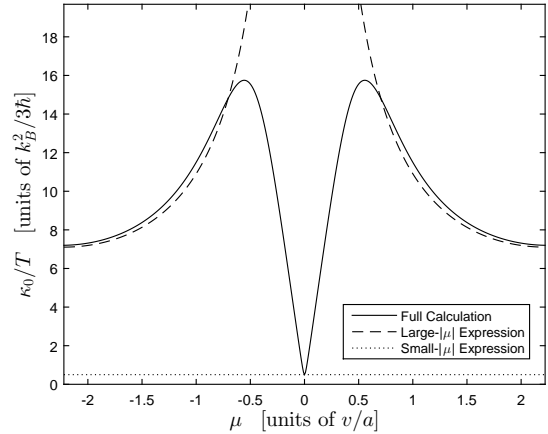}}}
\caption{Calculated universal-limit thermal conductivity, $\kappa_0/T$, as a function of chemical potential, $\mu$.  Solid curve denotes numerical solution of Eqs.~(\ref{eq:kappa0num}-\ref{eq:Deltanum}) for parameter values $\Delta_0=0.1 v/a$ and $\Gamma_0=0.01 v/a$.  Solution matches our large-$|\mu|$ expression (dashed) for $|\mu| \gg \Gamma_0$, then reaches a maximum at an intermediate value of $|\mu|$ before decreasing toward the value of our small-$|\mu|$ expression (dotted) for $|\mu| \ll \Gamma_0$.}
\label{fig:kappa0vsmu_limits}
\end{figure}

We would now like to look beyond the large-$|\mu|$ and small-$|\mu|$ limits and consider the transition between them by numerically evaluating Eqs.~(\ref{eq:kappa0c}-\ref{eq:Qk}) as a function of $\mu$.  This is easily done, but unlike the large and small $|\mu|$ limit calculations which were model independent (aside from the two conditions in Sec.~\ref{ssec:spectrum}), this calculation requires a model for $\Delta_k$, the proximity-induced superconducting order parameter of the TI-dSC interface state, and its results will necessarily depend (in the details) on that choice of model.  Since we are primarily interested in understanding the essential physics of this transition, without delving too deeply into the material-dependent details, we proceed by considering the following simple and rather standard expression for the order parameter of a generic $d$-wave superconductor
\begin{equation}
\Delta_k = \frac{\Delta_0}{2} \left( \cos k_xa - \cos k_ya \right)
\label{eq:Deltakmodel}
\end{equation}
which yields a gap velocity at ${\bf k}$ of the form
\begin{equation}
\vec{\bf v}_{\Delta k} \equiv \frac{\partial \Delta_k}{\partial {\bf k}} = \frac{\Delta_0 a}{2} \left( -\sin k_x a \,\hat{\bf x} + \sin k_y a \,\hat{\bf y} \right)
\label{eq:vDeltakmodel}
\end{equation}
Here, we have introduced two new model parameters, the gap maximum $\Delta_0$ and the lattice constant $a$.  Expressing all lengths in units of $a$ and all energies in units of $v/a$, we define dimensionless parameters $\tilde{\mu} \equiv \mu a/v$, $\tilde{\Gamma}_0 \equiv \Gamma_0 a/v$, and $\tilde{\Delta}_0 \equiv \Delta_0 a/v$, as well as a dimensionless wavevector with components $z_1 \equiv k_x a$ and $z_2 \equiv k_y a$.  Doing so, plugging Eqs.~(\ref{eq:Deltakmodel}-\ref{eq:vDeltakmodel}) into Eqs.~(\ref{eq:kappa0c}-\ref{eq:Qk}), and noting that all terms not proportional to the identity tensor integrate to zero, we find that the universal-limit thermal conductivity tensor reduces to a scalar and takes the convenient form
\begin{eqnarray}
\frac{\kappa_0}{T} &=& \frac{k_B^2}{3\hbar} \int \frac{d^2z}{8\pi} \biggl[ \left( L({\bf z}) + L(-{\bf z}) \right)^2 \nonumber \\
&+& \frac{\tilde{\Delta}_0^2}{2} \sin^2 z_1 \left( L({\bf z})^2 + L(-{\bf z})^2 \right) \biggr]
\label{eq:kappa0num}
\end{eqnarray}
where
\begin{equation}
L({\bf z}) \equiv \frac{\tilde{\Gamma}_0}{\tilde{\Gamma}_0^2 + (z-\tilde{\mu})^2 + \tilde{\Delta}({\bf z})^2}
\label{eq:Lnum}
\end{equation}
and
\begin{equation}
\tilde{\Delta}({\bf z}) \equiv \frac{\tilde{\Delta}_0}{2} \left( \cos z_1 - \cos z_2 \right)
\label{eq:Deltanum}
\end{equation}
The $k$-space integral is easily computed to obtain $\kappa_0/T$ as a function of $\mu$.  Results for $\Delta_0=0.1 v/a$ and $\Gamma_0=0.01 v/a$ are plotted in Fig.~\ref{fig:kappa0vsmu_limits} alongside the large and small $|\mu|$ limits.  (For the large-$|\mu|$ plot, we have used the model introduced in Eqs.~(\ref{eq:Deltakmodel}-\ref{eq:vDeltakmodel}) to obtain the nodal anisotropy ratio as a function of $\mu$, $v/v_\Delta=[(\tilde{\Delta}_0/\sqrt{2})\sin(\tilde{\mu}/\sqrt{2})]^{-1}$, and used that as input to Eq.~(\ref{eq:kappa0largemu}).)  Our numerical result matches the large-$\mu$ expression for $|\mu| \gg \Gamma_0$, peaking with decreasing $|\mu|$, before plunging down toward the small-$|\mu|$ value for $|\mu| \ll \Gamma_0$.

\begin{figure}
\centerline{\resizebox{3.25in}{!}{\includegraphics{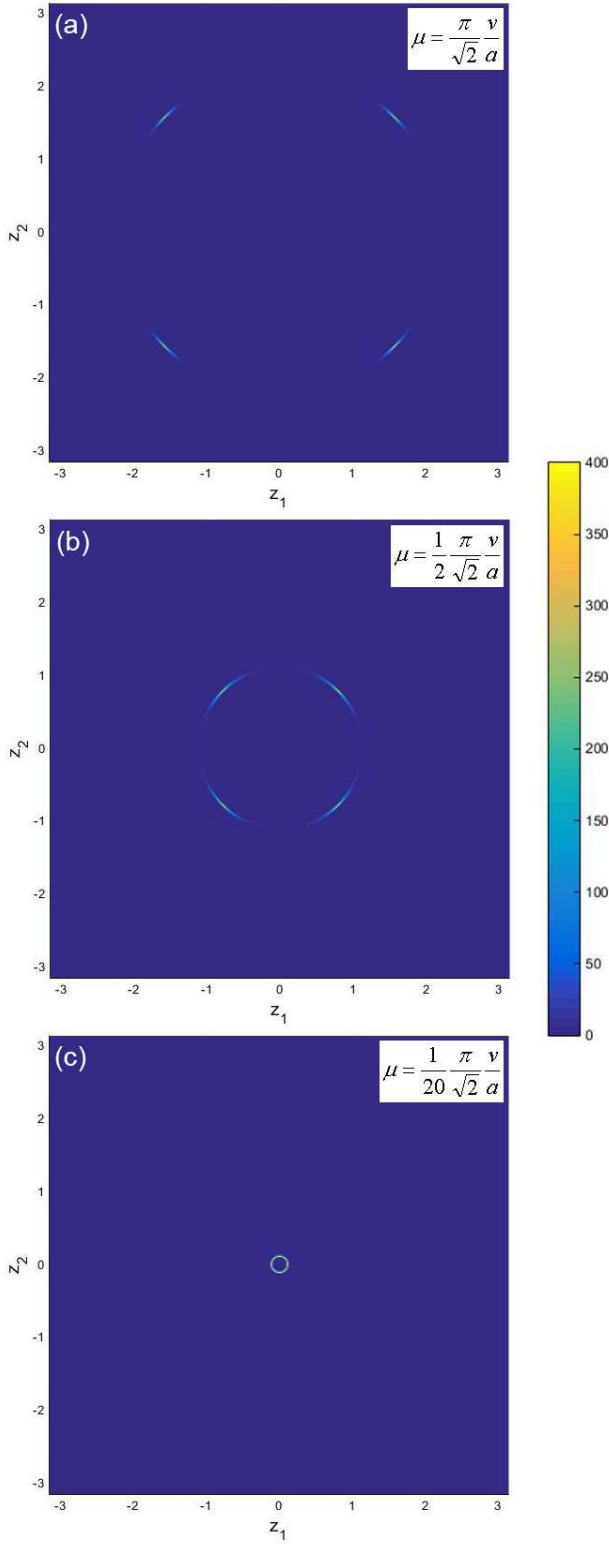}}}
\caption{Evolution of the $k$-space structure of the $\kappa_0/T$ integrand.  (a) [$\mu = \frac{\pi}{\sqrt{2}} \frac{v}{a}$]: Large-$|\mu|$ limit.  Four, well-separated, elliptical peaks within $\Gamma_0$ of the nodal points.  (b) [$\mu = \frac{1}{2} \frac{\pi}{\sqrt{2}} \frac{v}{a}$]: Nodal peaks closer to the origin, more anisotropic, and curving around the Fermi circle.  (c) [$\mu = \frac{1}{20} \frac{\pi}{\sqrt{2}} \frac{v}{a}$]: Nodal peaks have merged into an annular peak of width $\Gamma_0$.}
\label{fig:integ_nozoom}
\end{figure}

\begin{figure}
\centerline{\resizebox{3.25in}{!}{\includegraphics{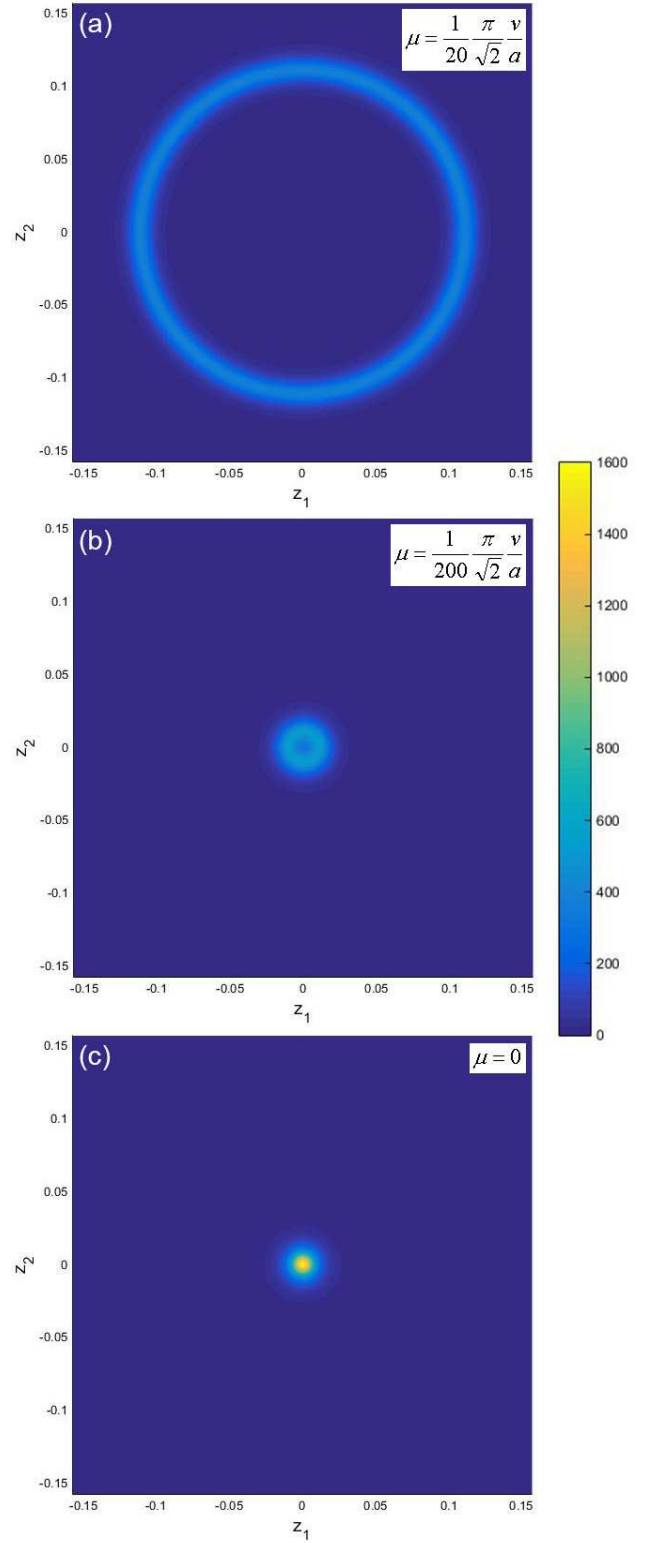}}}
\caption{Further evolution of the $k$-space structure of the $\kappa_0/T$ integrand, zoomed-in by a factor of 20.  (a) [$\mu = \frac{1}{20} \frac{\pi}{\sqrt{2}} \frac{v}{a}$]: Closeup view of the same annular peak shown in Fig.~\ref{fig:integ_nozoom}(c).  (b) [$\mu = \frac{1}{200} \frac{\pi}{\sqrt{2}} \frac{v}{a}$]: Width and radius of the annular peak now nearly equal.  (c) [$\mu = 0$] Small-$|\mu|$ limit.  Annular peak blurred into single, isotropic peak within $\Gamma_0$ of the origin.}
\label{fig:integ_zoom20}
\end{figure}

This behavior is best understood by considering the evolution of the $k$-space structure of the integrand of Eq.~(\ref{eq:kappa0num}) as a function of $\mu$, as shown for a series of $\mu$ values in Figs.~\ref{fig:integ_nozoom} and \ref{fig:integ_zoom20}.  The upper panel of Fig.~\ref{fig:integ_nozoom} illustrates the structure of the large-$\mu$ limit.  Here, for $\tilde{\mu}=\pi/\sqrt{2}$, the integrand is peaked within $\tilde{\Gamma}_0$ of four, well-separated, anisotropic nodal points.  Equal-intensity contours are (nearly) elliptical, squeezed in the direction parallel to the local Fermi surface.  In the middle panel, $\mu$ is reduced by a factor of 2, which draws the nodes closer to the origin.  The peaks are still well-separated, but less so than before, since the radius of the Fermi circle has decreased and the nodal anisotropy ratio has increased.  Thus, the peaks have begun to curve around the Fermi circle, toward each other, and the independent node approximation used to derive the large-$|\mu|$ expression of Eq.~(\ref{eq:kappa0largemu}) has begun to break down.  In the lower panel, $\mu$ is reduced by an additional factor of 10.  Now the independent node approximation has completely broken down, and the four anisotropic peaks have curved into each other, forming an annulus of width $\tilde{\Gamma}_0$ about the Fermi circle.  With decreasing $\mu$, the radius of this annular peak decreases, resulting in the decrease of $\kappa_0/T$ seen in Fig.~\ref{fig:kappa0vsmu_limits}.  We reproduce this image, zoomed-in about the annulus, in the upper panel of Fig.~\ref{fig:integ_nozoom}.  In the middle panel of that figure, $\mu$ is reduced by another factor of 10, such that it is nearly equal to $\Gamma_0$.  Now the width and radius of the annular peak are nearly equal.  As $\mu$ decreases further, the system is tuned toward the Dirac point inherited from the TI surface state and the isotropic node at the origin is revealed.  For $|\mu| \ll \Gamma_0$, the annular peak blurs into a single isotropic peak at the origin, of width $\tilde{\Gamma}_0$.  This is shown in the lower panel where $\mu=0$.  The integral over this single isotropic peak recovers the small-$|\mu|$ value of Eq.~(\ref{eq:kappa0smallmu}).  As $\mu$ becomes negative, the process reverses, dominated now by the $(-)$ branch of the quasiparticle excitation spectrum instead of the $(+)$ branch.  All else is the same, so $\kappa_0/T$ is even in $\mu$.

\begin{figure}
\centerline{\resizebox{3.25in}{!}{\includegraphics{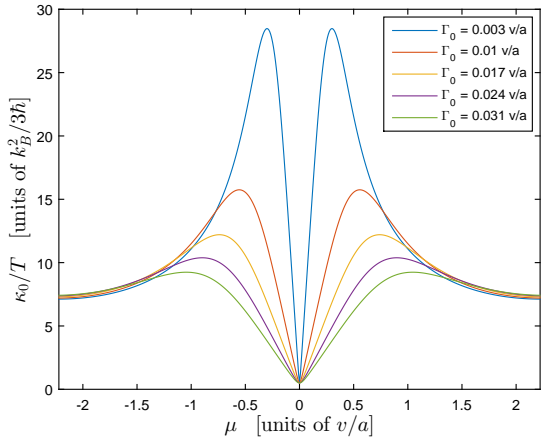}}}
\caption{Disorder dependence of calculated universal-limit thermal conductivity, $\kappa_0/T$, as a function of chemical potential, $\mu$.  We plot numerical solutions of Eqs.~(\ref{eq:kappa0num}-\ref{eq:Deltanum}) for $\Delta_0=0.1 v/a$ and five values of the impurity scattering rate, $\Gamma_0$.  Results are disorder-independent in both the large-$|\mu|$ and small-$|\mu|$ limits.  The transition between limits depends on disorder, with the peaks more prominent for smaller $\Gamma_0$, smoothing out with increasing disorder.}
\label{fig:kappa0vsmu_multipleG0}
\end{figure}

Results for five different values of the impurity scattering rate, $\Gamma_0$, are shown in Fig.~\ref{fig:kappa0vsmu_multipleG0}.  Note that in both the large-$|\mu|$ and small-$|\mu|$ limits, $\kappa_0/T$ is disorder-independent.  The transition between these limits does, however, depend on disorder, with the peaks of the $\kappa_0/T$ vs $\mu$ curves smoothed out for greater disorder.  This effect can be understood in terms of our integrand analysis (above).  As $|\mu|$ decreases from its largest values, increasing anisotropy ratio yields increasing $\kappa_0/T$ via our large-$|\mu|$ expression, Eq.~(\ref{eq:kappa0largemu}).  But for greater disorder, the independent node approximation that defines the large-$|\mu|$ limit breaks down sooner, as the four anisotropic peaks broaden with growing disorder and merge together earlier, limiting the enhancement of $\kappa_0/T$ with increasing anisotropy ratio.  The resulting annular peak is of greater width for greater disorder and therefore blurs into a single peak sooner, ushering in the small-$|\mu|$ limit as its radius becomes smaller than its width.

\section{Conclusions}
\label{sec:conclusions}
In this paper, we have calculated the universal-limit thermal conductivity, $\kappa_0$, as a function of chemical potential, $\mu$, due to quasiparticle excitations of the proximity-induced superconducting state at the 2D interface of a topological insulator and a $d$-wave superconductor.  In both the large-$|\mu|$ and small-$|\mu|$ limits, we have obtained simple closed-form expressions for $\kappa_0/T$, combined here from Eqs.~(\ref{eq:kappa0largemu}) and (\ref{eq:kappa0smallmu})
\begin{equation}
\frac{\kappa_0}{T} = \frac{k_B^2}{3\hbar} \left\{
\begin{array}{cc}
\frac{1}{2} \left( \frac{v}{v_\Delta} + \frac{v_\Delta}{v} \right) & \mbox{for} \;\; |\mu| \gg \Gamma_0 \\
\frac{1}{2} & \mbox{for} \;\; |\mu| \ll \Gamma_0
\end{array}
\right.
\label{eq:kappa0limits}
\end{equation}
where $v$ is the slope of the isotropic Dirac cone inherited from the TI surface state, $v/v_\Delta$ is the $\mu$-dependent anisotropy ratio of the four anisotropic Dirac cones of the proximity-induced $d$-wave superconducting state, and $\Gamma_0$ is the impurity scattering rate, the energy scale characterizing disorder in the system.  Note that the large-$|\mu|$ expression is exactly half the value obtained \cite{dur00} (per layer) for an ordinary $d$-wave superconductor: $\kappa_0^{dSC}/T = (k_B^2/3\hbar)(v_F/v_\Delta + v_\Delta/v_F)$.  This is an overt demonstration of the sense in which the underlying topological metal is ``half'' of an ordinary metal \cite{fu08}, and comes about because, for large $|\mu|$, only one of the two branches (positive or negative) of the isotropic Dirac cone contributes at a time.  For $|\mu| \ll \Gamma_0$, both branches contribute, but the four nodes have coalesced into one isotropic node at the origin of $k$-space.  Thus, the small-$|\mu|$ expression is equal to the standard dSC value (with anisotropy ratio equal to one), divided by four (since there is only one node instead of the usual four): $(1 + 1)/4 = 1/2$.  While $\kappa_0/T$ is disorder-independent in both of these limits, the transition between them, as a function of $\mu$, depends on disorder.  And furthermore, it depends, in the details, on the functional form of the proximity-induced order parameter, $\Delta_k$.  Adopting a simple model for $\Delta_k$ (Eq.~(\ref{eq:Deltakmodel})), we have calculated $\kappa_0/T$ across the full range of $\mu$, for different levels of disorder, as shown in Fig.~\ref{fig:kappa0vsmu_multipleG0}.  As $\mu$ decreases from its maximum value, the four nodal peaks of the integrand in Eq.~(\ref{eq:kappa0num}) become more anisotropic, resulting in an increase in $\kappa_0/T$, as per our large-$|\mu|$ expression.  But they also move closer together, eventually merging into an annular peak about the Fermi circle.  Along the way, the independent node approximation that defined the large-$|\mu|$ limit breaks down, and $\kappa_0/T$ reaches its maximum value, decreasing as $\mu$ decreases further and the Fermi circle shrinks.  Finally, as $\mu$ gets smaller than $\Gamma_0$, the annular peak blurs into an isotropic nodal peak at the origin, and $\kappa_0/T$ reaches its minimum at the value given by our small-$|\mu|$ expression.  As shown in Fig.~\ref{fig:kappa0vsmu_multipleG0}, the peaks in the $\kappa_0/T$ vs $\mu$ curve are more pronounced for smaller $\Gamma_0$, smoothing out with increasing disorder.

Note that we have assumed herein that the bulk band gap of the topological insulator extends well above and below the Dirac point of the surface state, such that $\mu$ could be varied over a wide range of energies without accessing the bulk valence or conduction bands.  In real materials, the available energy windows may be more restricted.  We have also assumed that the chemical potential can be accurately controlled, via gating, doping, or other means, and that proper contact can be made to the TI-dSC interface.  Both may present experimental challenges.

Our focus in this work has been on the evolution with changing chemical potential of the massless Dirac quasiparticle excitations of the TI-dSC interface state.  Results shed light on the essential features of low-temperature thermal transport due to these quasiparticles.  Further theoretical development, including incorporation of a subdominant spin-triplet order parameter, a more realistic disorder model, and vertex corrections to our diagrammatic calculation, are left for future work.

\begin{acknowledgments}
I am grateful to B. Burrington and G. C. Levine for very helpful discussions.  This work was supported by faculty startup funds provided by Hofstra University.
\end{acknowledgments}

\end{document}